# On sound insulation of pyramidal lattice sandwich structure

Jie Liu[1,†], Tingting Chen[2], Guilin Wen[1,2,†], Qixiang Qing[2], Ramin Sedaghati[3], Yi Min Xie[4,5]

[1]Center for Research on Leading Technology of Special Equipment, School of Mechanical and Electric Engineering, Guangzhou University, Guangzhou 510006, P. R. China

[2]State Key Laboratory of Advanced Design and Manufacturing for Vehicle Body, Hunan University, Changsha 410082, P. R. China

[3]Department of Mechanical, Industrial and Aerospace Engineering, Concordia University, Montreal, Quebec, Canada

[4]Centre for Innovative Structures and Materials, School of Engineering, RMIT University, Melbourne 3001, Australia

[5]XIE Archi-Structure Design (Shanghai) Co., Ltd., Shanghai 200092, China

†Correspondence and requests for materials should be addressed to J.L. (email: jliu@gzhu.edu.cn) or G.L.W. (email: glwen@gzhu.edu.cn)

**Abstract**

Pyramidal lattice sandwich structure (PLSS) exhibits high stiffness and strength-to-weight ratio which can be effectively utilized for designing light-weight load bearing structures for ranging from ground to aerospace vehicles. While these structures provide superior strength to weigh ratio, their sound insulation capacity has not been well understood. The aim of this study is to develop numerical and experimental methods to fundamentally investigate the sound insulation property of the pyramidal lattice sandwich structure with solid trusses (PLSSST). A finite element model has been developed to predict the sound transmission loss (STL) of PLSSST and simulation results have been compared with those obtained experimentally. Parametric studies is then performed using the validated finite element model to investigate the effect of different parameters in pyramidal lattice sandwich structure with hollow trusses (PLSSHT), revealing that the pitching angle, the uniform thickness and the length of the hollow truss and the lattice constant have considerable effects on the sound transmission loss. Finally a design optimization strategy has been formulated to optimize PLSSHT in order to maximize STL while meeting mechanical property requirements. It has been shown that STL of the optimal PLSSHT can be increased by almost 10% at the low-frequency band. The work reported here provides useful information for the noise reduction design of periodic lattice structures.

**Introduction**

Noise suppression is of critical importance with potential applications in various fields. Low frequency noise generated in ground and aerospace vehicles provide a harsh environment for passengers and crews and generally cause severe health issues under long-term exposure. Passive and active treatments have been widely used recently to control noise[1-12]. Compared with active methods[12], passive methods are simpler, cheaper, and more practical without requiring hardware and power sources associated with active systems. Structures with high stiffness and strength-to-weight ratio are generally inefficient sound barriers, thus designing light weight load-bearing structures with desirable acoustical properties is of paramount importance in aerospace ground and marine vehicles. Periodic lattice sandwich truss structures, comprising of two thin yet stiff face sheets separated by cores with open cell topology, namely, trusses, have recently emerged as potential candidates to design such structures[13-19]. Various open cell topological configurations have been proposed, including tetrahedron[13,17], 3D kagome[18], diamond[19], and pyramid[14], among which periodic lattice sandwich truss structure with pyramidal truss cores, namely PLSS, are considered to be very promising[13]. While properties of PLSS like out-of-plane compression and shear[20], free vibration[21,22], and shock loading[23] have been broadly exploited, very limited studies have been conducted on their sound insulation capability. Existing studies commonly use the equivalent or simplified model with one-dimensionally periodic unit cells to approximately analyze the sound insulation properties of PLSS[22-24]. It is therefore of significant importance to thoroughly explore the sound insulation capacity of the PLSS by developing a high fidelity three-dimensional numerical model and also designing experimental method to validate the model.

In this study, a high fidelity finite element (FE) model has been developed to investigate the sound transmission loss (STL) capability of PLSS with solid truss member (PLSSST). A proof-of-concept PLSSST has been fabricated and experimental setup has been designed to experimentally investigate its STL capability. It has been shown that good agreement exits between simulation and experimental results. Developed FE model has been utilized to simulate STL of PLSS with hollow truss members (PLSSHT) and to conduct parametric studies to realize the effects of various geometrical parameters on the STL of PLSSHT. Finally, a design optimization problem has been formulated to identify the optimal geometrical parameters of PLSSHT to maximize its STL.

## Results

**Geometric modelling of PLSSST and PLSSHT.** Figure 1A illustrates a typical PLSS, comprising of three parts, namely, the upper skin panel, the core, and the lower skin panel. The core, constituted by periodically arranged unit cells, is normally connected to the upper and lower panels by the adhesive. The unit cell, consisting of four inclined rods which can be solid (PLSSST) or hollow (PLSSHT). For PLSSST, the upper and lower panels have the exact same geometrical dimensions such that they have a length ($a$), the width ($b$), and the thickness ($d$). The azimuthal angle and the pitching angle between the rod and the lower panel are respectively $\alpha$ and $\beta$. The section area of the inclined solid rod is a rectangle with a width $w_1$ and a length $w_2$. The length of the rod is $L$. Besides, the distance between the adjacent inner vertex of each rod is $L_1$ and the outer base point of the each rod is at a distance $L_1/2$ from the outer boundary of the unit cell (see Figure 1B). The dimensions of the unit cell are $a_c = 2(L_1 + L\cos\beta\sin\alpha)$, $b_c = 2(L_1 + L\cos\beta\cos\alpha)$, and $h_c = L\sin\beta$. For PLSSHT, the unit cell is made of four inclined hollow rods, whose section area is regarded as a nested square, for simplicity, having a length of the inside square ($s$) and outside square ($w$), as shown in Figure 1C. Thus, the uniform thickness is $t = (w-s)/2$.

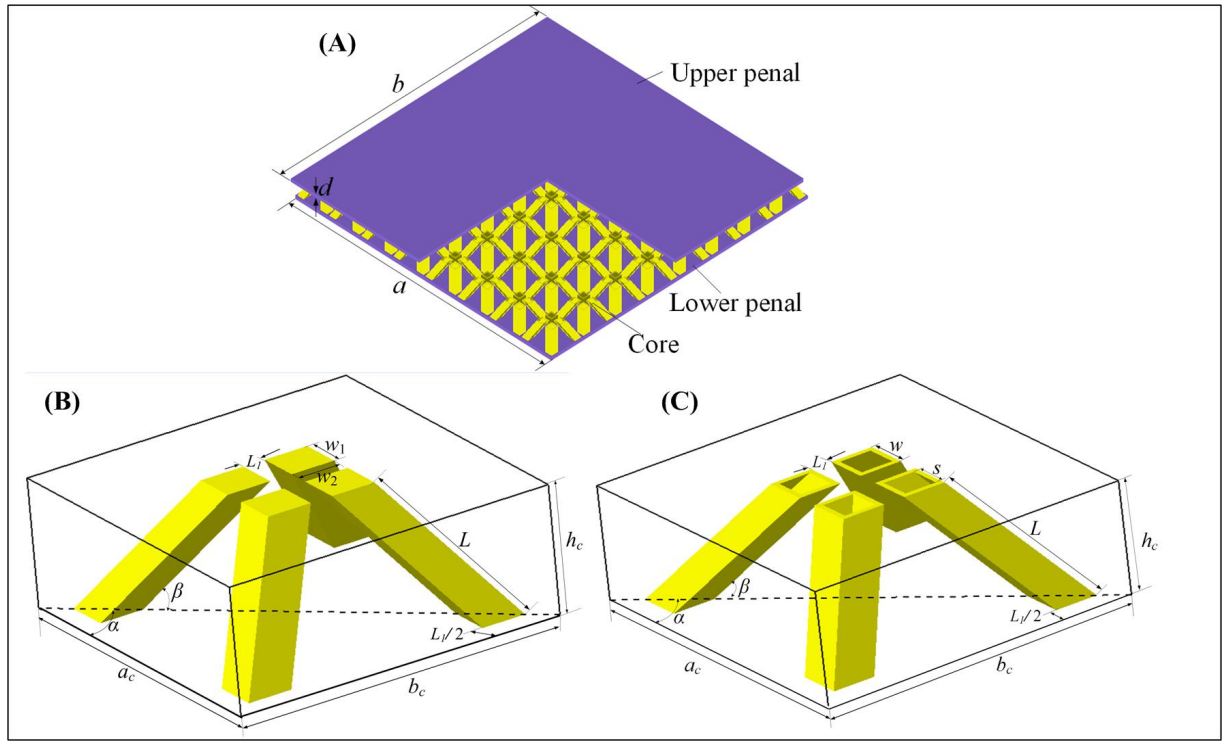

**Figure 1. Illustrations of PLSS.** (A) the geometry of PLSS whose core can be comprised by solid or hollow rods. (B) the unit cell of the PLSSST core with four inclined solid rods. (C) the unit cell of the PLSSHT core with four inclined hollow rods.

**Sound insulation of PLSSST.** Figure 2A depicts the sketch of a PLSST sound insulation panel in which the sound wave (incident wave) is incident to its upper panel of PLSSST, causing the panel to vibrate. The vibration is then transmitted to the lower panel via the truss core, generating radiated sound waves to the air fluid domain below the sandwich structure. Here a high fidelity finite element model has been developed to predict the sound transmission loss of PLSSST. An experimental setup has been designed to evaluate STL in wide frequency band and validate the FE model. In FE model, two 3D acoustic grids are established on both sides of the sandwich structure to simulate the reverberation chamber and the anechoic chamber respectively (see Figure 2B). The detailed setup of the FE model can be found in Supplementary Information, Section "Numerical Simulation ".

Figure 2C shows the fabricating process of the real PLSSST. For convenience, the fabrication for one unit cell has been presented. Firstly, we use ethoxyline resin to glue four trusses to the thin panel by carefully placing their location and wait until the ethoxyline resin is cured. Then, we flip over the structure from the previous procedure and adhere another thin panel to the structure. Last, we flip over the structure again and wait until the ethoxyline resin is properly cured. It should be pointed out that all the trusses should be glued to the thin panel before flipping over the structure in the real fabrication process. Left side of Figure 2D presents the real PLSSST with the panels and trusses all made of aluminium and right side of Figure 2D displays the sound insulation cavity made from five PLSSSTs used to conduct the experiments.

Figure 2E depicts the comparison results of the STL from the simulation and three experiments (the detailed experimental setup can be found in Supplementary Information, Section "Experimental Details"). As it can be realized, good agreement exists between simulation and experimental results except for the low-frequency band, roughly below 100 Hz. This is mainly due to the difficulty in accurate measurement of STL in the low-frequency range.

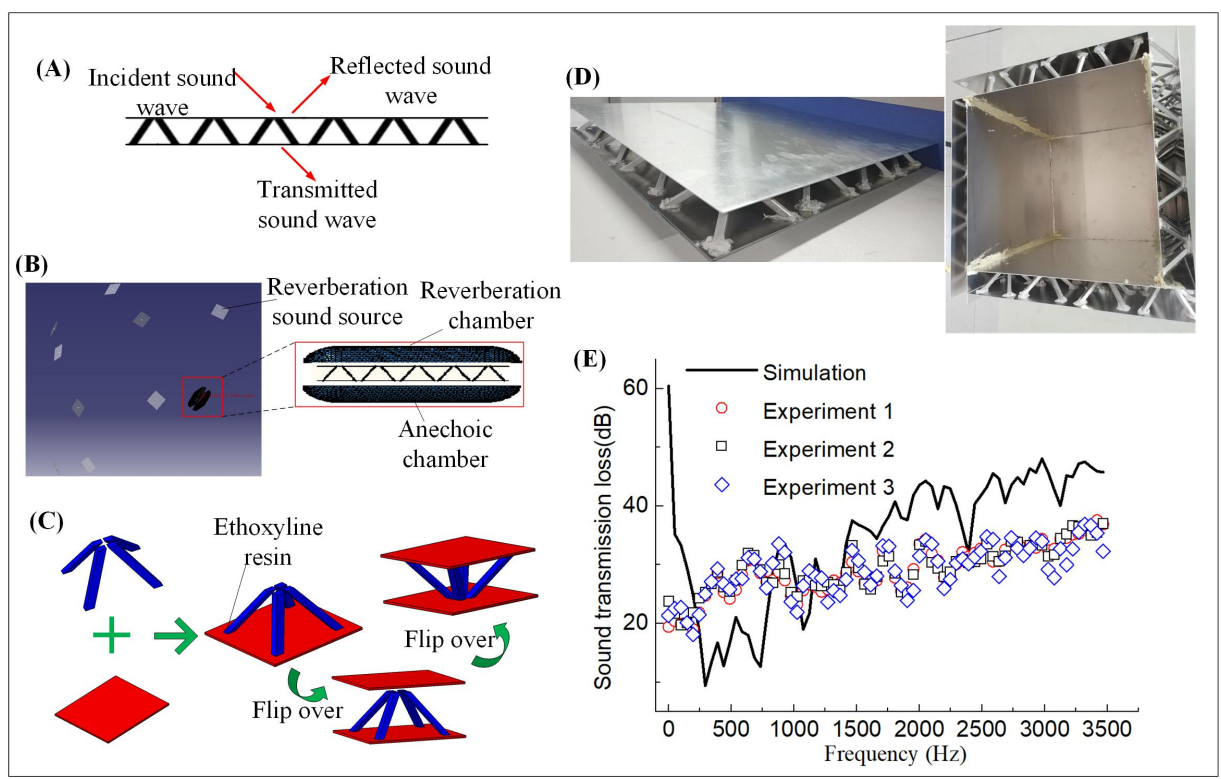

**Figure 2.** (A) Diagram for the sound transmission of PLSS. (B) FE model for the sound insulation of PLSS. (C) schematic drawing for the flowchart of fabricating PLSS. (D) the real PLSSST (left) and the sound insulation cavity made from PLSSST (right). (E) comparison results of the sound

transmission loss of simulation and experiments.

**STL analysis of Parametric Study of PLSSHT.** To reveal the STL capacity of PLSSHT, parametric numerical analysis is performed at the frequency band of 100-5000Hz by using the developed verified FE model. Figure 3A-D show the influences of the pitching angel ($β$), the uniform thickness ($t$) and the length of the hollow truss ($L$), and the lattice constant ($Ln$) on the STL of PLSSHT, demonstrating that these key factors all considerably affect the STL. Figure 3A shows that for frequency nearly below 1500 Hz, the STL initially decreases and then increases as the value of the pitching angle from 45° to 75°. At frequency of approximately 3000 Hz, there appears a trough when $β$ equals to 60° whereas crests emerge when $β$ equals to 45° or 75°. This situation is exactly the opposite at frequency of around 4000 Hz. It can also be realized that the fundamental natural frequency occurs respectively at 1756.21 Hz, 1405.22 Hz, and 1927.65 Hz for $β$ equals to 45°, 60°, and 75°. Results demonstrate that structural optimization can be performed if one wants to maximize the STL of PLSSHT at specific frequency band, i.e. 2500 Hz-4500 Hz, when $β$ is chosen as the design parameter. Figure 3B indicates that the STL increases slightly by changing the value of $t$ from 0.25 mm to 1.0 mm when the frequency is smaller than 1700 Hz. Particularly, every 0.25 mm increase in $t$ enlargers the STL by around 1.3 dB. It is worth pointing out that there are circumstances that STL is smaller with a larger $t$, for example, when frequency is near 2000 Hz or 4000 Hz. The fundamental natural frequency is respectively 1658 Hz, 1755 Hz, and 17561 Hz when $t$ equals to 0.25 mm, 0.5 mm, and 1.0 mm. It can be found from Figure 3C that STL becomes larger by increasing the value of $L$ in almost whole frequency band, revealing that it may be an very effective way to improve the STL of PLSSHT. Figure 3D displays that STL decreases by almost 9 dB at least when the value of $Ln$ increase from 15 mm to 40 mm at frequency band 2500 Hz-3200 Hz, demonstrating that $Ln$ significantly influences the STL of PLSSHT. Although we identify that these four key factors influence the STL of PLSSHT in various degrees, we have to adjust their values to meet the demand of STL in actual applications.

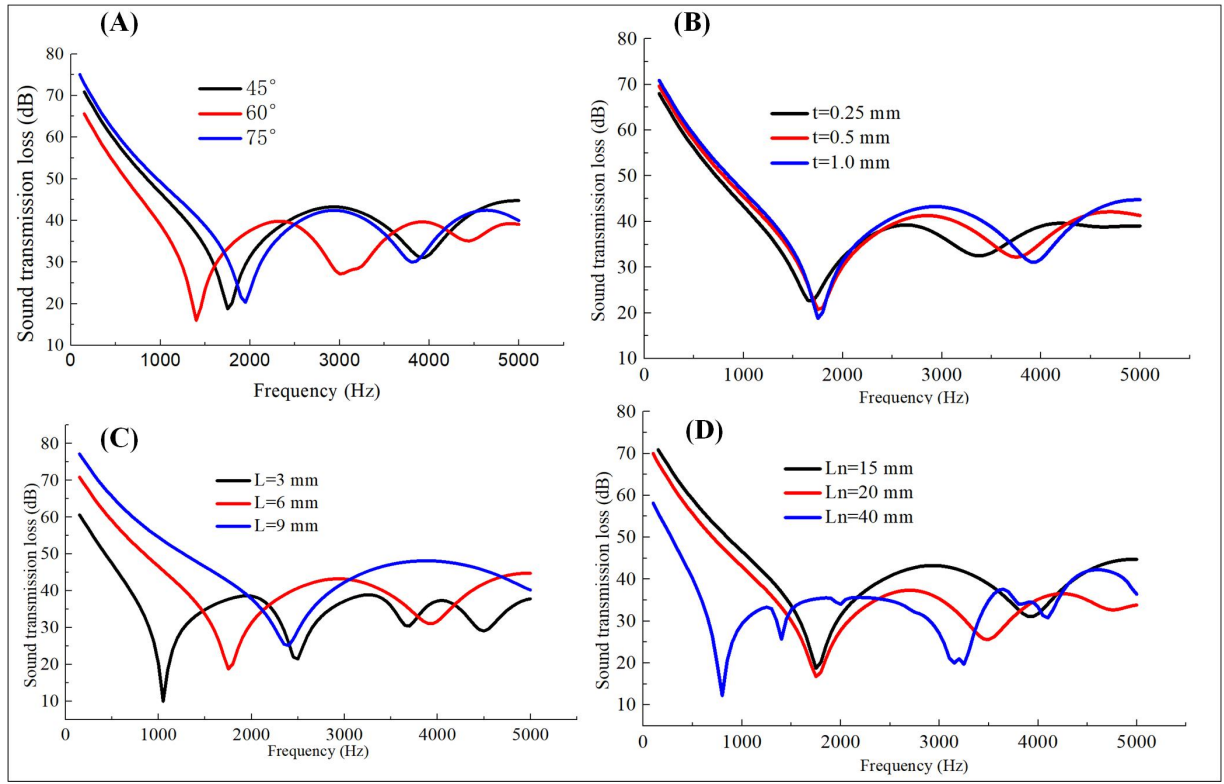

**Figure 3. Effects of the geometric parameters of the unit cell on STL.** (A) pitching angle of the hollow truss. (B) uniform thickness of the hollow truss. (C) length of the hollow truss. (D) lattice constant.

**Structural Design Optimization of PLSSHT.** To further enhance the STL capacity of PLSSHT at the concerned frequency band, a design optimization problem has been formulated to search for the optimal parameters to maximize STL. It should be noticed that two inequalities are introduced to the optimization model (see equations S7 and S8). Besides, the total mass of the PLSSHT is kept unchanged, which can be regarded as the equality constraint (see equation S9). The four key factors, namely, the pitching angel ($\beta$), the uniform thickness ($t$), the length of the inside square ($s$), and the length of the hollow truss ($L$) are chosen as the design parameters. The details of the definitions of structural optimization design can be found in Supplementary Information, Section "Optimization". The optimal value for design parameters are found to be $s$=0.47 mm, $L$=45.5 mm, $t$=0.5 mm, and $\beta$=58.6° after solving GA (Compared results can be also found in Table S3). The results for STL of the optimized PLSSHT for both low frequency and high frequency bandwidth are shown in Figure 4 and compared with those of initial design. As it can be realized the STL of the optimized PLSSHT has been improved by almost 5 dB at the low-frequency band, i.e. 125-500 Hz as shown in Figure 4A. It is noted that the noise sources in the cabin of the operating aircraft are mainly caused by the superposition of random broadband and harmonic noise, whose frequency are normally within this range. Similarly, the STL of the optimized PLSSHT has also been considerably magnified at the high-frequency band, i.e. 2500-4000 Hz as shown in Figure 4B.

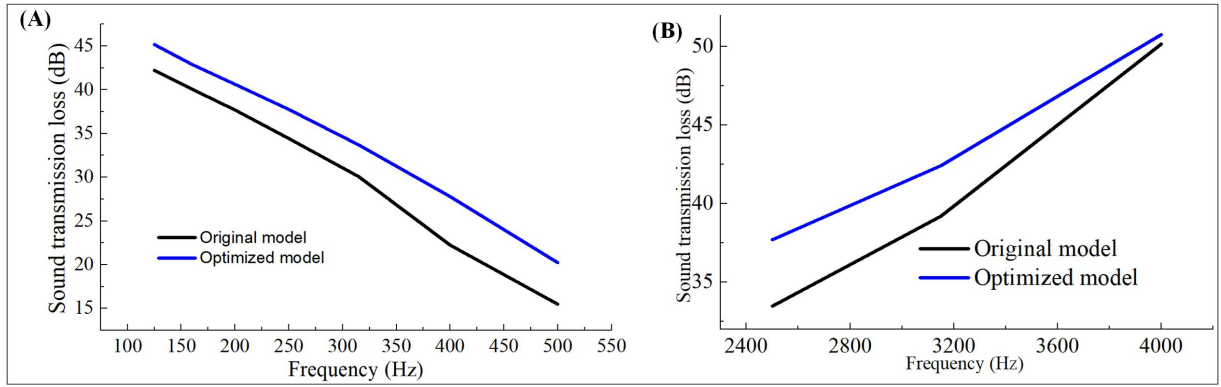

**Figure 4. Comparison results of the optimized model and original model.** (A) low-frequency band, 125-500 Hz. (B) high-frequency band, 2500-4000 Hz.

**Discussion**

In this study, the STL capacity of PLSSST has been studied both numerically and experimentally. A high fidelity FE model has been developed to predict the STL behavior of PLSSST under wide range of frequencies. To validate the developed FE model, a novel experimental set up has been designed to measure STL of a sound insulation cavity made of PLSSST panels. It is shown that relatively good agreement exists between simulation and experimental results. The the validate FE model has been utilized to predict the STL of PLSSHT panels followed by through parametric studies to investigate the effect of key geometrical parameters of PLSSHT on the STL. Finally, to find the optimal PLSSHT with maximum STL, a structural optimization problem has been formulated under mechanical and volume constraints. The STL of the optimal PLSSHT has been evidently increased, particularly at the low-frequency band. Although the proposed size optimization methodology can improve the STL ability of PLSSHT to a certain extent, it may not be still able to meet the requirement of the noise reduction of actual applications. To this end, continuum topology optimization technique[25-28] could be a promising candidate with strong capacities for significantly improving the structural functionalities. Considering this, future work will focus on maximizing the STL of PLSSHT by introducing holes to the trusses of the unit cells, while reducing the weigh via topology optimization methods. Moreover, it is worth noting that sound insulation with the aid of semi-active method[29-31] by using smart materials, i.e. magnetorheological fluids, has been actively investigated as this method has combined the merits of the passive method and active method. Another research will concentrate on the semi-active sound insulation method based on the origami/kirigami-based sandwich structures or metamaterials[32-36] in the future.

## Methods

**Numerical simulations.** Numerical simulations are conducted by the finite element method based on commercial software LMS Virtual.Lab 11. The design of experiment method and polynomial response surface method are used to explicitly express the relationship between the STL and the key design parameters. Genetic Algorithm is employed to solve the optimization problem.

**Acoustic measurements.** Four microphones (378B02, PCB Piezotronics, America), whose sensitivities are all 50 mV/Pa, are used to measure the sound from the sound insulation cavity. The white noise signal is amplified by the wireless loudspeaker, and the dynamic sound pressure signal of the measuring point is collected in real time by the data acquisition system composing the embedded controller (NI PXI-1042Q) and the dynamic data acquisition card (NI PXI-4472). The sound pressure signal in the time domain is converted to the frequency domain via Fourier Transform and the spectrum analysis is performed in LMS Test. Lab to get the A-weighted frequency-sound pressure level curve.

## Acknowledgements

This research was financially supported by the National Natural Science Foundation of China (No. 11672104).

## Author Contributions

J.L. and G.L.W. conceived the key ideas and supervised the whole project. J.L. and T.T.C. performed the numerical simulations. T.T.C. and Q.X.Q. conducted the experiments. G.L.W., R.S. and Y.M.X. contributed to discussions. J.L. drafted the manuscript. G.L.W., R.S. and Y.M.X. revised the manuscript.

## Additional Information

**Competing Interests:** The authors declare no competing financial interests.

Supplementary Information
for
# On sound insulation of pyramidal lattice sandwich structure

By *Jie Liu\*, Tingting Chen, Guilin Wen\*, Qixiang Qing, Ramin Sedaghati, Yi Min Xie*

[*] Prof. G. Wen, A/Prof. J. Lie

Center for Research on Leading Technology of Special Equipment, School of Mechanical and

Electric Engineering

Guangzhou University, Guangzhou 510006, P. R. China

E-mail: glwen@gzhu.edu.cn & jliu@gzhu.edu.cn

[*] Prof. G. Wen, A/Prof. Q. Qing, T. Chen

State Key Laboratory of Advanced Design and Manufacturing for Vehicle Body

Hunan University, Changsha 410082, P. R. China

[*] Prof. Ramin Sedaghati

Department of Mechanical, Industrial and Aerospace Engineering
 Concordia University, Montreal, Quebec, Canada

[*] Prof. Yi Min Xie

Centre for Innovative Structures and Materials, School of Engineering,

RMIT University, Melbourne 3001, Australia

& XIE Archi-Structure Design (Shanghai) Co., Ltd., Shanghai 200092, China

**Numerical Simulation**
The FE model of the PLSSST has been developed in commercial software LMS Virtual.Lab 11. Direct acoustic vibration coupling theory available in the LMS Virtual.Lab is utilized to study the STL behavior of PLSSST. The three-dimensional (3D) models of PLSSST, the reverberation chamber, and the anechoic chamber are, first, built in CATIA (the related geometric parameters are summarized in Table S1) and saved as Initial Graphics Exchange Specification (IGES) files, and then they are imported into Altair HyperMesh for meshing. The element type for the panel and the rod are Pshell and Psolid, respectively; while the reverberation chamber and the anechoic chamber are occupied by Tetrahedral elements. PLSSST is modeled by 421,024 elements with a nominal maximum size of 1 mm and 466,130 nodes.The material of PLSSST is aluminium with the material density, Young's modulus, and Poisson ration as 2810 Kg/mm$^3$, 71 GPa, and 0.33, respectively. The loss factor of the panel and the bar are set to be 0.0001 and 0.4, respectively, considering that viscoelastic materials would be used to bond the bar and the panel.
In FE model, two acoustic grids are established on both sides of the sandwich structure to simulate the reverberation chamber and the anechoic chamber. The reverberation chamber and the anechoic chamber are divided into 802,332 elements with a nominal maximum size of 4 mm and 164,468 nodes by using 3D acoustic meshes. Both of them are assigned to air property with the sound velocity and mass density being 340 m/s and 1.225 Kg/m$^3$ respectively. To ensure that the reverberation room completely radiates sound power and the anechoic room totally absorbs sound power, both of the two acoustic meshes are assigned to an automatic matching feature (AML), at the same time, the outer surface of the anechoic chamber is endowed a radiation surface property. The inner surfaces of the reverberation chamber and the anechoic chamber are respectively coupled with the upper and lower panels of the PLSSST, and defined as the acoustic-structure coupling surface for energy transfer. Sound source excitation is a theoretical uniform reverb sound source formed by 12 plane waves with a sound pressure amplitude of 1 Pa. The AML property imparted on the reverberation surface makes the reverb sound power radiate completely in the reverberation room, and the AML attribute conferred on the outdoor surface of the sound damper. And the properties of the radiation surface used for far-field calculations make the anechoic chamber infinitely characteristic. Acoustic waves enter the anechoic chamber through the measuring part and finally radiate out in the anechoic chamber in an infinitely distant manner. After simply supporting the constraints on the four sides of the PLSSST, the coupled vibration response of the model is obtained, and finally the STL curve of the PLSSST is acquired.

Table S1 Geometric parameters of PLSSST

| Parameter | $a$ (mm) | $b$ (mm) | $\alpha$ (°) | $\beta$ (°) | $d$ (mm) | $L$ (mm) | $L_1$ (mm) | $w_1$ (mm) | $w_2$ (mm) |
|---|---|---|---|---|---|---|---|---|---|
| Value | 360 | 360 | 45 | 45 | 1 | 54 | 8 | 6 | 8.48 |

**Experimental details**

The experiments are conducted in a large and quiet room in order to extremely reduce the possible influence of the background noise on the test results. Four test points, marked A[#], B[#], C[#], and D[#], are respectively arranged at the front, rear, left, and right sides of the sound source with their position being 1 m above the ground and 1.5 m away from the sound source (see Fig. S1A). A wireless speaker, which is covered by the sound insulation cavity (see Fig. S1B), is placed in the middle of the desk to generate the incident sound wave, which is regarded as the sound source (see Fig. S1A). High precision microphones are laid at the test points to measure the incident sound wave (see Fig. S1B). Professional data acquisition system acquires the dynamic sound pressure signals at measuring points (see Fig. S1C). This dynamic sound pressure signal is further analyzed using FFT analyzer to gain the signal in the frequency domain, and then spectrum analysis is performed to obtain the A-weighted sound pressure level. The difference of the mean value of sound pressure level at the four test points before and after the sound insulation cavity is regarded as the insertion loss. It should be noted that here we use the insertion loss to verify the STL from the numerical simulation.

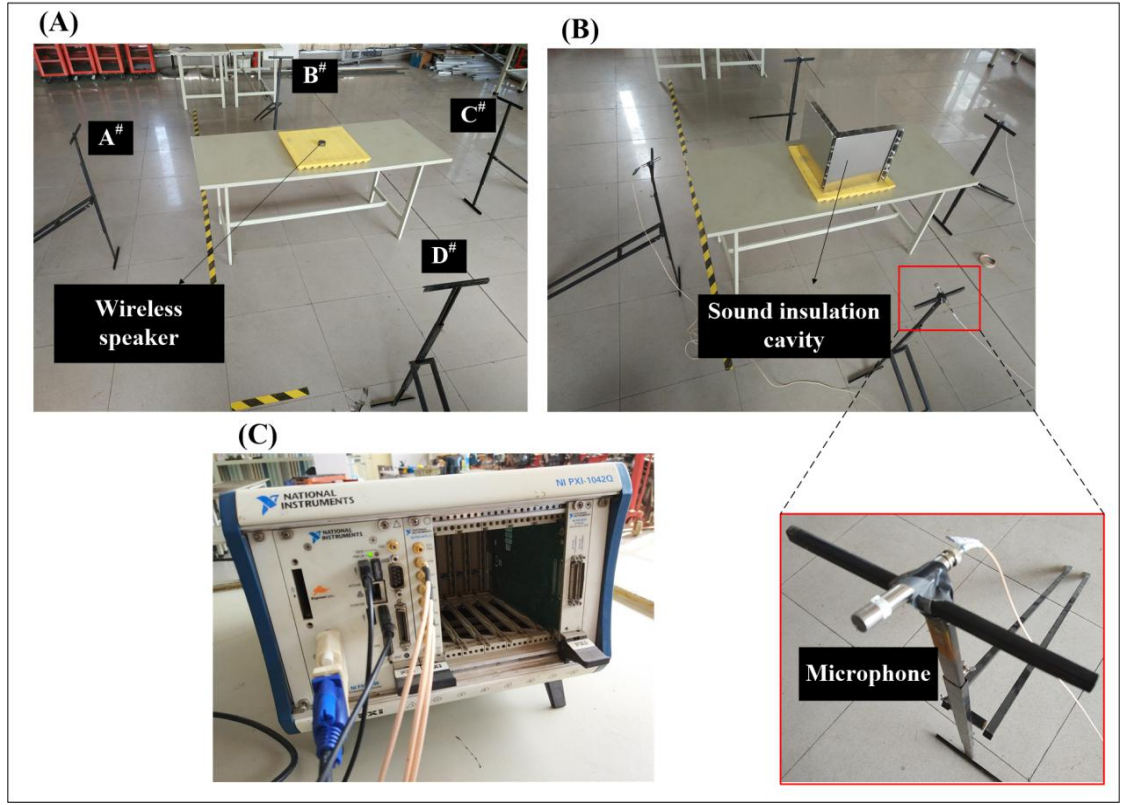

**Fig. S1. Experimental setup**. (A) A wireless speaker (sound source) to produce incident noise. (B) Covering the sound source by using the sound insulation cavity made of PLSS and placing four microphones around 1.5 meters from the sound insulation cavity to receive the STL and convert sound signals into electrical signals. (C) data acquisition system.

**Optimization**

A. **Response surface model of STL**

Conducting optimization on the full finite element model is computationally expensive and also may not render accurate optimal solution due to possible noisy behavior of the response. Thus, in this research study design of experiment (DoE) combined with response surface method (RSM) has been utilized to generate smooth response function which relates STL to identified key design parameters. It should be noted that here the out-of-plane mechanical properties of the PLSSHT are considered. When PLSSHT is subjected to out-of-plane compressive load, the failure of the sandwich rod including its buckling and breaking is the main factor that causes the failure of the overall structure which should be considered as the constraints in the optimization. We first consider the buckling of the sandwich rod as the constraint. To achieve this, the following inequality is given,

$$\lambda \geq \lambda_c \qquad (S1)$$

where $\lambda$ and $\lambda_c$ are the slenderness and the critical slenderness of the sandwich rod, respectively.

The critical slenderness can be expressed as,

$$\lambda_r = \sqrt{\pi^2 E / \sigma_p} \tag{S2}$$

where $E$ represents the Young's modulus of the constitutive material and $\sigma_p$ is the proportional limit stress.

The slenderness of the sandwich rod can be written as,

$$\lambda = \frac{L}{\sqrt{(w^2 + s^2)/3}} \tag{S3}$$

To purely investigate the STL property of PLSSHT, we keep the weight unchanged. Since only one constitutive material is used, keeping the weight constant is equivalent to making all the PLSSHTs have the same unit cell volume, which can be formulated as,

$$V_c = 4(w^2 - s^2)L + 2d(2L_1 + \sqrt{2}L\cos\beta)^2 = V_c^0 \tag{S4}$$

Where the initial unit cell volume, $V_c^0$, based on the initial parameters is 5257.8 mm3. The three key factors, namely, the pitching angel ($\beta$), the length of the inside square ($s$), and the length of the hollow truss ($L$) are chosen as the independent design parameters. However, considering equality constraint in Eq. (6) and after submitting the value for w and $V_c^0$, one can obtain the pitching angle with respect to design parameters $s$ and $L$ and then substitute them in Eqs. (4) and (5), thus eliminating the equality constraint and reducing the number of design variable to two, $s$ and $L$. Design-of-Experiment (DoE) based on Latin Sampling method[1] is utilized to generate 22 sample points to in the specified design space. Table S2 provides the identified sampling points using DoE and the evaluated response STL on these points.

Table S2 The generated DoE sample points and associated STL

| No. | $L$ (mm) | $t$ (mm) | $\beta$ (°) | $s$ (mm) | STL (dB) |
|---|---|---|---|---|---|
| 1 | 56.589 | 0.260 | 64.965 | 0.950 | 31.821 |
| 2 | 51.215 | 0.350 | 62.288 | 0.769 | 32.682 |
| 3 | 49.714 | 0.313 | 61.221 | 0.843 | 32.571 |
| 4 | 46.357 | 0.440 | 59.161 | 0.591 | 33.196 |
| 5 | 55.650 | 0.344 | 64.756 | 0.783 | 32.446 |
| 6 | 51.602 | 0.303 | 62.383 | 0.864 | 32.433 |
| 7 | 48.779 | 0.201 | 60.192 | 1.068 | 31.735 |
| 8 | 50.777 | 0.379 | 62.095 | 0.712 | 32.825 |
| 9 | 47.767 | 0.461 | 60.262 | 0.547 | 33.293 |
| 10 | 50.162 | 0.339 | 61.592 | 0.792 | 32.690 |
| 11 | 53.370 | 0.227 | 63.154 | 1.016 | 31.859 |
| 12 | 52.155 | 0.360 | 62.883 | 0.749 | 32.651 |
| 13 | 54.434 | 0.215 | 63.696 | 1.039 | 31.723 |

| 14 | 55.438 | 0.470 | 64.949 | 0.529 | 32.823 |
| 15 | 47.405 | 0.448 | 59.970 | 0.573 | 33.241 |
| 16 | 45.207 | 0.238 | 57.633 | 0.993 | 31.941 |
| 17 | 49.463 | 0.282 | 60.955 | 0.905 | 32.376 |
| 18 | 48.089 | 0.406 | 60.367 | 0.658 | 33.063 |
| 19 | 46.537 | 0.240 | 58.704 | 0.989 | 32.072 |
| 20 | 52.754 | 0.271 | 62.957 | 0.929 | 32.171 |
| 21 | 53.673 | 0.410 | 63.879 | 0.650 | 32.746 |
| 22 | 54.555 | 0.427 | 64.396 | 0.616 | 32.745 |

Based on the data in Table S2 and using response surface method (RSM) based on the second-order polynomial function[2], the following response function has been derived:

$$\text{STL}^{RMS} = 23.06101 - 2.54701s + 0.496559L - 2.4012s^2 + 0.07298sL - 0.00583L^2 \quad (S5)$$

It is worthy to point out that the coefficient of multiple determination ($R^2$), root mean square error (*RMSE*), and the adjusted coefficient of multiple determination ($R^2_{adj}$) of the developed RSM are function are found to be 0.9953, 0.0402, and 0.9938, demonstrating that it is capable of preceding the response STL accurately.

**B. Design variables**

We constrain the design variables to lie in the intervals $s \in (0.47\text{mm}, 1.07\text{ mm})$, $L \in (45\text{mm}, 60\text{ mm})$, $t \in (0.2\text{mm}, 0.5\text{ mm})$, and $\beta \in (25°, 75°)$. It should be reminded that the fluctuations of the design variables are consistent with those shown in section A.

**C. Constraints**

To avoid the breaking of the sandwich rod when PLSSHT is under the out-of-plane compressive load, the crushing strength of the sandwich rod, $\sigma_{pk}$, should be greater than that of the reference model, $\sigma_{pk0}$, as:

$$\sigma_{pk} = \frac{\pi E(w^2 + s^2)}{3L^2} \frac{2(w^2 - s^2)}{\sin\beta(L\cos\beta + \sqrt{2}L_1^2)} \left( \sin^2\beta + \frac{w^2 + s^2}{L^2}\cos\beta \right) \geq \sigma_{pk}^0 \quad (S6)$$

Note that $\sigma_{pk0}$ corresponds to the reference model whose geometric parameters can be found in Table S3. With carefully investigated, no breaking is found for the reference model. It is worth pointing out that conservative outcomes may be obtained for the optimization problem by using equation (S6).

Table S3 Geometric parameters of reference model (RM) and optimal model (OM)

| Parameter | $a$ (mm) | $b$ (mm) | $\alpha$ (°) | $\beta$ (°) | $d$ (mm) | $L$ (mm) | $L_1$ (mm) | $s$ (mm) | $w$ (mm) | $t$ (mm) |
|---|---|---|---|---|---|---|---|---|---|---|
| RM | 360 | 360 | 45 | 60 | 1 | 48 | 8 | 0.87 | 1.47 | 0.3 |
| OM | 360 | 360 | 45 | 58.6 | 1 | 45.5 | 8 | 0.47 | 1.47 | 0.5 |

Thus, we have the following eleven constraints, as

$$g_1(s,L,t) = \lambda_c - \lambda \leq 0 \tag{S7}$$

$$g_2(s,L,t,\beta) = \sigma_{pk}^0 - \sigma_{pk} \leq 0 \tag{S8}$$

$$g_3(s) = -s + 0.47 \leq 0 \tag{S9}$$

$$g_4(s) = s - 1.07 \leq 0 \tag{S10}$$

$$g_5(L) = -L + 45 \leq 0 \tag{S11}$$

$$g_6(L) = L - 60 \leq 0 \tag{S12}$$

$$g_7(t) = -t + 0.2 \leq 0 \tag{S13}$$

$$g_8(t) = t - 0.5 \leq 0 \tag{S14}$$

$$g_9(\beta) = -\beta + 25° \leq 0 \tag{S15}$$

$$g_{10}(\beta) = \beta - 75° \leq 0 \tag{S16}$$

$$h_1(s,L,t,\beta) = V_c - V_c^0 = 0 \tag{S17}$$

### D. Objective

We use the RSM in equation (S5) to replace the real STL of PLSSHT to serve as the objective function, as

$$f = \text{STL}^{RSM} \tag{S10}$$

Hence, the optimization problem can be mathematically formulated as,

$$\begin{aligned} \text{Min}: \quad & -f \\ \text{s.t.}: \quad & g_1(s,L,t) \leq 0 \\ & g_2(s,L,t,\beta) \leq 0 \\ & h_1(s,L,t,\beta) = 0 \end{aligned} \tag{S11}$$

The optimization problem stated in Eq. (S11) can be solved efficiently. Here GA[3] has been utilized as the optimizer to solve the formulated optimization problem due to its proven capability to catch global optimum solution. In the GA, the maximum iteration number, the probability of genetic variation, the crossover probability, and the convergence error setting as 200, 0.05, 0.6, and 0.5%, respectively.